\documentclass[doublecol]{epl2} 

\usepackage{amsmath}

\title{Quantum diffusion due to scattering non-locality in nanoscale semiconductors}
\shorttitle{Quantum diffusion due to scattering non-locality} 

\author{Roberto Rosati \and Fausto Rossi}
\shortauthor{R. Rosati \etal}

\institute{                    
Department of Applied Science and Technology, Politecnico di Torino \\
C.so Duca degli Abruzzi 24, 10129 Torino, Italy
}

\pacs{72.10.-d}{general formulation of electronic transport theory}
\pacs{73.63.-b}{electronic transport in nanoscale materials and structures}
\pacs{85.35.-p}{nanoelectronic devices}

\abstract{
In view of its local character, the semiclassical or Boltzmann theory is intrinsically unable to describe transport phenomena on ultrashort space and time scales, and to this purpose genuine quantum-transport approaches are imperative.
By employing a density-matrix simulation strategy recently proposed, we shall demonstrate its power and flexibility in describing quantum-diffusion phenomena in nanoscale semiconductors. In particular, as for the case of carrier-carrier relaxation in photoexcited semiconductors, our analysis will show the failure of simplified dephasing models in describing phonon-induced scattering non-locality, pointing out that such limitation is particularly severe for the case of quasielastic dissipation processes.
}

\begin{document}

\maketitle

In spite of the intrinsic validity limits of the semiclassical transport theory \cite{b-Jacoboni89}, during the last decades a number of Boltzmann-like Monte Carlo simulation schemes have been successfully employed for the investigation of new-generation semiconductor nanodevices \cite{Ryzhii98,Iotti01,Koehler02,Callebaut04,Lu06,Bellotti08,Jirauschek10,Matyas10,Iotti10,Matyas13}. 
Such modeling strategies ---based on the neglect of carrier phase coherence--- are however unable to properly describe space-dependent ultrafast phenomena. To this aim, the crucial step is to adopt a quantum-mechanical description of the carrier subsystem \cite{b-Datta05,b-Haug04,b-Haug07,b-Jacoboni10,b-Rossi11}; this can be performed at different levels, ranging from phenomenological dissipation/decoherence models \cite{Gmachl01,Iotti05a} to quantum-kinetic treatments \cite{Rossi02,Axt04}.
Indeed, in order to overcome the intrinsic limitations of the semiclassical picture in properly describing ultrafast space-dependent phenomena ---e.g., real-space transfer and escape versus capture processes--- Jacoboni and co-workers have proposed a quantum Monte Carlo technique \cite{Brunetti94}, while Kuhn and co-workers have proposed properly designed quantum-kinetic treatments \cite{Reiter07}; 
however, due to their high computational costs, these non-Markovian density-matrix approaches are currently unsuitable for the design and optimization of new-generation nanodevices.

In order to overcome such limitations, a conceptually simple as well as physically reliable quantum-mechanical generalization of the conventional Boltzmann theory has been recently proposed \cite{Rosati13b}. The latter preserves the power and flexibility of the semiclassical picture in describing a large variety of scattering mechanisms; indeed, employing a microscopic derivation of generalized scattering rates based on a recent reformulation of the Markov limit \cite{Taj09}, a density-matrix equation has been derived, able to properly account for space-dependent ultrafast dynamics in semiconductor nanostructures.

Primary goal of this Letter is to apply the density-matrix approach proposed in \cite{Rosati13b} to the investigation of genuine quantum-diffusion phenomena in state-of-the-art nanoscale semiconductors. 
Indeed, while the numerical analysis presented in \cite{Rosati13b} can be regarded as a ``proof of concept'' and is limited to a qualitative evaluation of the initial charge-density variation in a GaAs structure due to acoustic-phonon scattering (whose impact in a real nanodevice is definitely negligible), in this Letter we shall present space- as well as time-resolved simulated experiments of ultrafast carrier dynamics in GaN-based nanomaterials in the presence of a strong electron-LO phonon coupling, providing a quantitative investigation of free-carrier versus scattering-induced diffusion. 
Moreover, as for the case of carrier-carrier relaxation in photoexcited semiconductors \cite{Rossi02}, our analysis will show the failure of simplified dephasing models in describing phonon-induced scattering non-locality, pointing out that such limitation is particularly severe for the case of quasielastic dissipation processes.

Generally speaking, the key interplay between phase coherence and dissipation/decoherence phenomena may be conveniently described through the equation of motion for the electronic single-particle density matrix \cite{b-Rossi11}
\begin{equation}\label{SBE}
\frac{\upd \rho_{\alpha_1\alpha_2}}{\upd t} = 
\left.\frac{\upd \rho_{\alpha_1\alpha_2}}{\upd t}\right|_{\rm sp} +
\left.\frac{\upd \rho_{\alpha_1\alpha_2}}{\upd t}\right|_{\rm mb}\ ,
\end{equation}
where
\begin{equation}\label{sp}
\left.\frac{\upd \rho_{\alpha_1\alpha_2}}{\upd t}\right|_{\rm sp} =
\frac{\epsilon_{\alpha_1} - \epsilon_{\alpha_2}}{i \hbar} \rho_{\alpha_1\alpha_2} 
\end{equation}
accounts for the coherent evolution dictated by the non-interacting single-particle Hamiltonian (here $\alpha$ and $\epsilon_\alpha$ denote the single-particle eigenstates and energy levels corresponding to the nanodevice potential profile), while the last (many-body) term encodes dissipation and decoherence processes, arising from the energy exchange between the carriers and the host material.
Equation~(\ref{SBE}) applies to a broad variety of problems ranging from quantum-transport to ultrafast-optics phenomena, a remarkable example being the semiconductor Bloch equations \cite{b-Haug04}.

It is worth stressing that the degree of accuracy of the density-matrix equation (\ref{SBE}) is intimately related to an appropriate choice of its last term. Indeed, oversimplified treatments ---e.g., purely phenomenological $T_1 T_2$ models as well as conventional Markov treatments \cite{Iotti05b} may lead to a violation of the positive-definite character of the density matrix $\rho_{\alpha_1\alpha_2}$, a mandatory prerequisite of any quantum-mechanical time evolution fulfilled, e.g., by employing so-called Lindblad superoperators \cite{Lindblad76}.

The simplest parameter-free form of the many-body contribution in (\ref{SBE}) is given by the following relaxation-time model \cite{Dolcini13}:
\begin{equation}\label{RTA}
\left.\frac{\upd \rho_{\alpha_1\alpha_2}}{\upd t}\right|_{\rm mb}
=
-\,\frac{\Gamma_{\alpha_1} + \Gamma_{\alpha_2}}{2}
\left(\rho_{\alpha_1\alpha_2} - \rho^\circ_{\alpha_1\alpha_2}\right)\ .
\end{equation}
Here $\rho^\circ_{\alpha_1\alpha_2} = f^\circ_{\alpha_1} \delta_{\alpha_1\alpha_2}$ is the equilibrium density matrix dictated by the host material, and
\begin{equation}\label{Gamma}
\Gamma_\alpha = \sum_s \sum_{\alpha'} P^s_{\alpha'\alpha}
\end{equation}
is the total scattering rate (i.e., summed over all final states $\alpha'$ and relevant interaction mechanisms $s$) corresponding to the microscopic transition probabilities $P^s_{\alpha'\alpha}$ of the semiclassical transport theory given by the standard Fermi's golden rule \cite{b-Jacoboni89}.
Within such relaxation-time paradigm, the diagonal contributions ($\alpha_1 = \alpha_2$) describe population transfer (and thus energy dissipation) toward the equilibrium carrier distribution $f^\circ_{\alpha_1}$ according to the relaxation rate $\Gamma_{\alpha_1}$, whereas the off-diagonal contributions ($\alpha_1 \ne \alpha_2$) describe a decay of the inter-state polarizations according to the decoherence/dephasing rate 
$(\Gamma_{\alpha_1} + \Gamma_{\alpha_2})/2$.

In spite of its simple structure and straightforward physical interpretation, the relaxation-time term (\ref{RTA}) is intrinsically different from the in- minus out-scattering structure of the Boltzmann collision term, and for this reason it may lead to a significant overestimation of decoherence processes (see below). To overcome this serious limitation, inspired by the conventional semiclassical theory, a novel Markov procedure has recently been   proposed \cite{Taj09}. Following this alternative adiabatic-decoupling scheme, it is possible to perform a microscopic derivation of a Lindblad-like scattering superoperator; more specifically, for any one-electron interaction mechanism $s$ (carrier-phonon, carrier-plasmon, carrier-impurity scattering, etc.) in the low-density limit one gets
\begin{equation}\label{Lindblad}
\left.\frac{\upd \rho_{\alpha_1\alpha_2}}{\upd t}\right|_{\rm mb}
=
\left.\frac{\upd \rho_{\alpha_1\alpha_2}}{\upd t}\right|_{\rm in}
-
\left.\frac{\upd \rho_{\alpha_1\alpha_2}}{\upd t}\right|_{\rm out}\ ,
\end{equation}
where
\begin{equation}\label{in}
\left.\frac{\upd \rho_{\alpha_1\alpha_2}}{\upd t}\right|_{\rm in}
=
\frac{1}{2} \sum_s \sum_{\alpha'_1\alpha'_2} 
\mathcal{P}^s_{\alpha_1\alpha_2,\alpha'_1\alpha'_2}
\rho_{\alpha'_1\alpha'_2}\,
+ \textrm{H.c.}
\end{equation}
and
\begin{equation}\label{out}
\left.\frac{\upd \rho_{\alpha_1\alpha_2}}{\upd t}\right|_{\rm out}
=
\frac{1}{2} \sum_s \sum_{\alpha'_1\alpha'_2} 
\mathcal{P}^{s *}_{\alpha'_1\alpha'_1,\alpha_1\alpha'_2}
\rho_{\alpha'_2\alpha_2}\,+ \textrm{H.c.}
\end{equation}
(H.c. denoting the Hermitian conjugate)
are so-called in- and out-scattering superoperators \cite{Taj09} written in terms of generalized scattering rates
\begin{equation}\label{calP}
\mathcal{P}^s_{\alpha_1\alpha_2,\alpha'_1\alpha'_2} =
A^s_{\alpha_1\alpha'_1} A^{s *}_{\alpha_2\alpha'_2} \ ,
\end{equation}
where the explicit form of the Lindblad matrix elements $A^s_{\alpha\alpha'}$ depends on the particular interaction mechanism considered.
It is vital to stress that the diagonal elements of the generalized scattering rates in (\ref{calP}) coincide with the conventional scattering rates of the semiclassical theory, namely $\mathcal{P}^s_{\alpha'\alpha',\alpha\alpha} = P^s_{\alpha'\alpha}$.

The quantum-mechanical approach proposed in \cite{Rosati13b} (and employed in the present work) is then given by the density-matrix equation (\ref{SBE}) equipped with the microscopic scattering superoperator in (\ref{Lindblad}).
Given the (time-dependent) solution $\rho_{\alpha_1\alpha_2}$ of our density-matrix equation, any single-particle quantity can be straightforwardly derived. In particular, for the case of a one-dimensional system with coordinate $z$ and momentum $k$, the space and momentum charge distributions are simply given by
\begin{equation}\label{nz}
n(z) = \sum_{\alpha_1\alpha_2} \phi^{ }_{\alpha_1}(z) \rho_{\alpha_1\alpha_2} \phi^*_{\alpha_2}(z)
\end{equation}
and
\begin{equation}\label{nk}
n(k) = \sum_{\alpha_1\alpha_2} \tilde{\phi}^{ }_{\alpha_1}(k) \rho_{\alpha_1\alpha_2} \tilde{\phi}^*_{\alpha_2}(k)\ ,
\end{equation}
where $\phi_\alpha(z) \equiv \langle z \vert \alpha \rangle$ denotes the real-space wavefunction  corresponding to the eigenstate $\alpha$, and $\tilde{\phi}_\alpha(k) \equiv \langle k \vert \alpha \rangle$ its Fourier transform.

Combining the prescription in (\ref{nz}) with the density-matrix equation (\ref{SBE}), the time evolution of the spatial carrier density can be written as
\begin{equation}\label{dndt-tot}
\frac{\partial n(z)}{\partial t} 
=
\left.\frac{\partial n(z)}{\partial t}\right|_{\rm sp}
+
\left.\frac{\partial n(z)}{\partial t}\right|_{\rm mb}
\end{equation}
with
\begin{equation}\label{dndt-sp}
\left.\frac{\partial n(z)}{\partial t}\right|_{\rm sp}
=
\sum_{\alpha_1\alpha_2} 
\phi^{ }_{\alpha_1}(z) 
\frac{\epsilon_{\alpha_1}-\epsilon_{\alpha_2}}{i \hbar}
\rho_{\alpha_1\alpha_2} 
\phi^*_{\alpha_2}(z)
\end{equation}
and
\begin{equation}\label{dndt-mb}
\left.\frac{\partial n(z)}{\partial t}\right|_{\rm mb} = 
\sum_{\alpha_1\alpha_2} \phi^{ }_{\alpha_1}(z) 
\left.\frac{\upd \rho_{\alpha_1\alpha_2}}{\upd t}\right|_{\rm mb}
\phi^*_{\alpha_2}(z)\ .
\end{equation}
It is possible to show that, also for the simplest case of a semiconductor bulk system (whose single-particle states are momentum eigenstates: $\alpha =
k$), in the presence of a non-parabolic band the single-particle evolution in (\ref{dndt-sp}) deviates from the classical diffusion term \cite{Hess96,Demeio05,Rosati13a}. 
Moreover, opposite to the Boltzmann theory, the scattering-induced variation in (\ref{dndt-mb}) is in general different from zero (see Fig.~\ref{Fig3} below); indeed, the action of the relaxation-time term (\ref{RTA}) as well as of the quantum-mechanical scattering superoperator (\ref{Lindblad}) is spatially non-local, in clear contrast to the Boltzmann collision term. Such non-local character will also manifest itself via a scattering-induced current, which can still be described via a generalized charge continuity equation; indeed, as pointed out in \cite{Gebauer04}, simplified dissipation/decoherence models (e.g. effective Pauli master equations or relaxation-time treatments) may lead to charge-continuity violations; This is definitely not the case of the present Lindblad-like microscopic approach.

Let us now apply the density-matrix formalism just described to the time-dependent simulation of ultrafast carrier dynamics in nanoscale materials. To this aim, we consider an effective one-dimensional GaN-based nanostructure, whose main energy-dissipation/decoherence mechanism is carrier-LO phonon scattering. In order to mimic the main features of a realistic GaN-based material, the following parameters have been employed:
effective mass $m^* = 0.2 m_\circ$ ($m_\circ$ denoting the free-electron one), LO-phonon energy $\epsilon_{\rm LO} = 80$\,meV, and average bulk carrier-LO phonon scattering rate $\tau_{\rm LO} = 25$\,fs. 

For all the simulated experiments presented below we have chosen as initial condition a single-particle density matrix $\overline{\rho}_{\alpha_1\alpha_2}$ corresponding to a gaussian carrier distribution both in space and momentum, namely
\begin{equation}\label{Gauss}
\overline{n}(z) \propto \frac{e^{-\frac{z^2}{2 \overline{\Delta}_z^2}}}{\sqrt{2\pi} \,\overline{\Delta}_z}
\ , \qquad
\overline{n}(k) \propto \frac{e^{-\frac{k^2}{2 \overline{\Delta}_k^2}}}{\sqrt{2\pi}\, \overline{\Delta}_k}\ ,
\end{equation}
where $\overline{\Delta}_z$ describes the degree of spatial localization of our initial state, and $\overline{\Delta}_k = \frac{\sqrt{m^*k_B T}}{\hbar}$ 
describes the thermal fluctuations of our carrier gas.
It is easy to show that this corresponds to choosing an initial density matrix of the form
\begin{equation}\label{rhocirc}
\overline{\rho}_{\alpha_1\alpha_2} \propto 
\int \upd z\, \upd k 
{\cal W}^*_{\alpha_1\alpha_2}(z,k)
\frac{
e^{-\frac{z^2}{2 \overline{\Delta}_z^2}}
e^{-\frac{k^2}{2 \overline{\Delta}_k^2}}
}
{
\sqrt{2\pi}\, \overline{\Delta}_z \overline{\Delta}_k
}
\ ,
\end{equation}
where
\begin{equation}\label{calW}
{\cal W}^{ }_{\alpha_1\alpha_2}(z,k) = \int \upd z'
\phi^{ }_{\alpha_1}\left(z+\frac{z'}{2}\right) 
\frac{e^{-i k z'}}{\sqrt{2\pi}} \phi^*_{\alpha_2}\left(z-\frac{z'}{2}\right)
\end{equation}
is the Weyl-Wigner transform corresponding to our single-particle basis states $\alpha$ \cite{b-Rossi11}. 

\begin{figure}
\onefigure{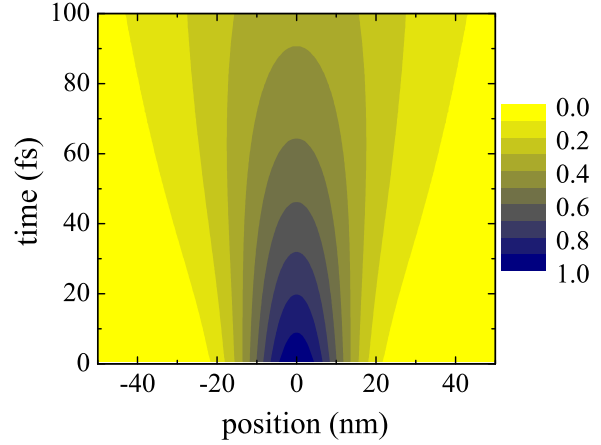}
\caption{
Room-temperature quantum-diffusion dynamics in a GaN bulk system obtained via the Lindblad scattering superoperator in (\ref{Lindblad}): sub-picosecond time evolution of the spatial carrier density corresponding to the initial mixed state in (\ref{rhocirc}) with $\overline{\Delta}_z = 10$\,nm (see text).
}
\label{Fig1}
\end{figure}

Our first set of room-temperature simulated experiments corresponds to a GaN bulk system (i.e., no confinement potential profile). 
Figure \ref{Fig1} displays the sub-picosecond time evolution of the spatial carrier density corresponding to the initial mixed state in (\ref{rhocirc}) with $\overline{\Delta}_z = 10$\,nm, obtained employing the Lindblad scattering superoperator in (\ref{Lindblad}). 
Such ultrafast diffusion dynamics is the result of a highly non-trivial interplay between the single-particle and many-body contributions in (\ref{dndt-tot});
indeed, it is well known that also in the presence of a spatially local (i.e., Boltzmann) scattering model (for which the contribution in (\ref{dndt-mb}) is always equal to zero) any  scattering-induced carrier redistribution tends to speed up the diffusion process \cite{b-Ashcroft11}.
In order to better evaluate the genuine diffusion contribution due to scattering non-locality, it is then crucial to start our simulated experiments from a thermalized carrier distribution; this has been realized adopting the initial state in (\ref{rhocirc});
indeed, for a parabolic-band bulk system in the absence of scattering non-locality, the time evolution of the spatial carrier density is described by the following (time-dependent) Gaussian distribution
\begin{equation}\label{nzt}
n(z,t) \propto \frac{e^{-\frac{z^2}{2 \Delta_z^2(t)}}}{\sqrt{2\pi} \Delta_z(t)}
\end{equation}
with
\begin{equation}\label{Deltazt}
\Delta_z(t) = \overline{\Delta}_z \sqrt{1+\frac{t^2}{\tau_d^2}}\ ,
\end{equation}
where 
$\tau_d = (m^* \overline{\Delta}_z)/(\hbar\overline{\Delta}_k)$ 
describes the typical time scale of the scattering-free diffusion process (for the case of Fig.~\ref{Fig1} the latter is about $70$\,fs).

\begin{figure}
\onefigure{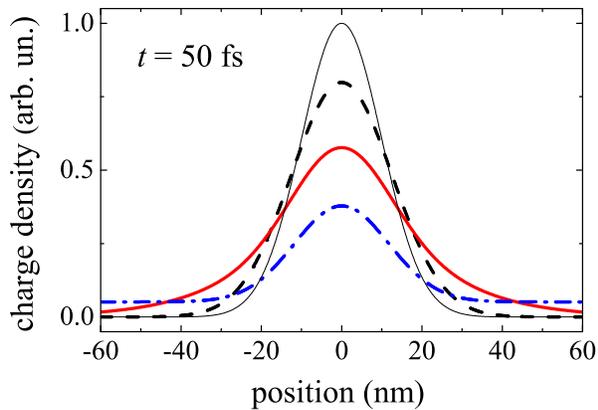}
\caption{
Room-temperature spatial carrier density at $t = 50$\,fs in a GaN bulk system corresponding to the initial mixed state in (\ref{rhocirc}) with $\overline{\Delta}_z = 10$\,nm. 
Here, the local-scattering solution in (\ref{nzt}) (dashed curve) is compared
to the corresponding results obtained adopting as scattering models the Lindblad superoperator (\ref{Lindblad}) (solid curve) as well as the relaxation-time model (\ref{RTA}) (dash-dotted curve); the initial distribution (thin solid curve) is also shown (see text).
}
\label{Fig2}
\end{figure}

In order to better elucidate the impact of scattering-induced diffusion, in Fig.~\ref{Fig2} the local-scattering solution in (\ref{nzt}) at $t = 50$\,fs (dashed curve) is compared
to the corresponding results obtained adopting as scattering models the Lindblad superoperator (\ref{Lindblad}) (solid curve) as well as the relaxation-time model (\ref{RTA}) (dash-dotted curve).
As we can see, both scattering treatments give rise to a diffusion speed up, and the effect is particularly pronounced for the case of the relaxation-time model (see dash-dotted curve).

\begin{figure}
\onefigure{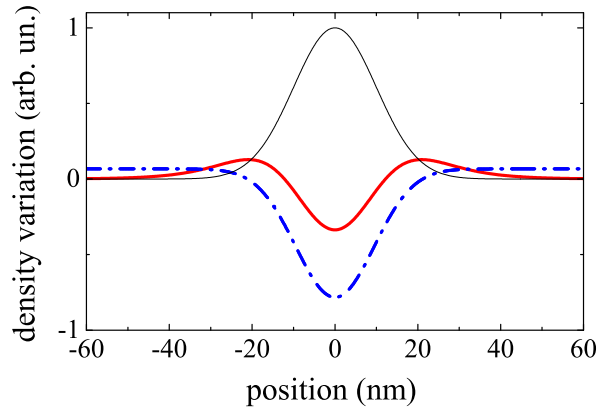}
\caption{
Phonon-induced spatial non-locality in a GaN bulk system:
time derivative of the spatial carrier density (see Eq.~(\ref{dndt-tot})) at $t = 0$ 
corresponding to the Lindblad superoperator (\ref{Lindblad}) (solid curve) and  to the relaxation-time model (\ref{RTA}) (dash-dotted curve); the initial distribution (thin solid curve) is also shown (see text).
}
\label{Fig3}       
\end{figure}

To better identify the physical origin and relative magnitude of such diffusion speed up, it is useful to analyze the shape of the scattering-induced density variation in (\ref{dndt-mb}). Figure \ref{Fig3} shows
the time derivative of the spatial carrier density (see Eq.~(\ref{dndt-tot})) at $t = 0$ 
corresponding to the Lindblad superoperator (\ref{Lindblad}) (solid curve) and  to the relaxation-time model (\ref{RTA}) (dash-dotted curve). 
As we can see, within the Lindblad-superoperator model (solid curve) the phonon-induced time variation displays a negative peak corresponding to a replica of the initial distribution (so-called out-scattering contribution induced by the last term in (\ref{Lindblad})) and, more importantly, a positive contribution extending over a much larger range (so-called in-scattering contribution induced by the first term in (\ref{Lindblad})).
This is exactly the signature of scattering-induced spatial non-locality we were looking for.
Also for the case of the simplified relaxation-time model (dash-dotted curve) we deal with a significant scattering non-locality; however, opposite to the Lindblad model, in this case the action of the scattering comes out to be totally non-local, as confirmed by its nearly constant values at large coordinate values.
Such highly non physical behavior gives rise to an increased dissipation/decoherence dynamics, which in turn results in the significant overestimation of the diffusion process reported in Fig.~\ref{Fig3}.
 It is worth noting a qualitative similarity between the scattering-induced diffusion scenario of Fig.~\ref{Fig3} and the numerical results presented in \cite{Gebauer04}; we stress, however, that the transport model employed in their numerical analysis is not of Lindblad form and, more importantly, neglects non-diagonal density-matrix elements, giving rise to a possible overestimation of the diffusion process.

\begin{figure}
\onefigure{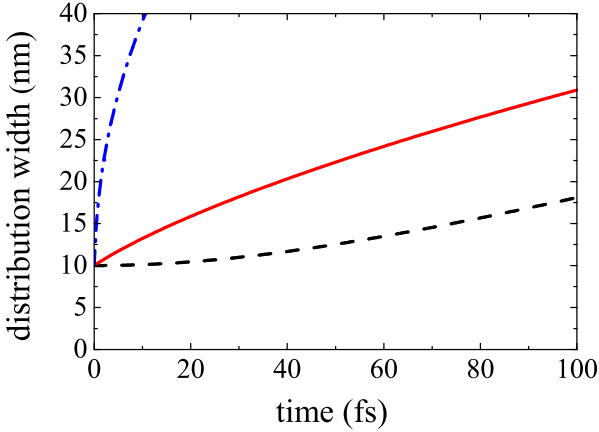}
\caption{
Effective spatial-distribution width $\lambda$ in (\ref{lambda}) as a function of time.
Here, the local-scattering result (see Eq.~(\ref{nzt})) (dashed curve) is compared
to the corresponding results obtained adopting as scattering models the Lindblad superoperator (\ref{Lindblad}) (solid curve) as well as the relaxation-time model (\ref{RTA}) (dash-dotted curve) (see text).
}
\label{Fig4}       
\end{figure}

To quantify the amount of extra diffusion reported in Figs.~\ref{Fig2} and \ref{Fig3}, let us introduce the effective carrier distribution width
\begin{equation}\label{lambda}
\lambda = 
\sqrt{
\frac{
\int z^2 n(z) \upd z
}{
\int n(z) \upd z
}
}\ .
\end{equation}
Figure \ref{Fig4} shows the time evolution of the above effective distribution width $\lambda$.
Here, the local-scattering result $\lambda = \Delta_z(t)$ (dashed curve) is compared
to the corresponding results obtained adopting as scattering models the Lindblad superoperator (\ref{Lindblad}) (solid curve) as well as the relaxation-time model (\ref{RTA}) (dash-dotted curve).
As expected, the relaxation-time model gives rise to a strong overestimation of the diffusion process (see dash-dotted curve) compared to the Lindblad-superoperator treatment (solid curve).

As anticipated, the relaxation-time model in (\ref{RTA}) does not exhibit the well established in- minus out-scattering structure of the Boltzmann collision term as well as of the Lindblad superoperator in (\ref{Lindblad}); it follows that within such simplified model the decay of the inter-state polarization is not dictated by a balance between in- and out-contributions, but is determined by out-scattering contributions only, leading to an overestimation of decoherence.
In order to elucidate this crucial point, let us start by analyzing the explicit form of Eq.~(\ref{SBE}) for the case of the relaxation-time model in (\ref{RTA}). By denoting with
\begin{equation}\label{rhoi}
\rho_{\alpha_1\alpha_2}^{\rm i}(t) = \rho^{ }_{\alpha_1\alpha_2}(t) e^{-\frac{(\epsilon_{\alpha_1}-\epsilon_{\alpha_2}) t}{i\hbar}}
\end{equation}
the single-particle density matrix written in the interaction picture, the time evolution of its non-diagonal ($\alpha_1 \ne \alpha_2$) elements is given by
\begin{equation}\label{SBERTA}
\frac{\upd \rho_{\alpha_1\alpha_2}^{\rm i}}{\upd t}
=
-\,\frac{\Gamma_{\alpha_1} + \Gamma_{\alpha_2}}{2}\,\rho_{\alpha_1\alpha_2}^{\rm i}\ , 
\end{equation}
which shows that, in addition to the free rotation in (\ref{rhoi}), the inter-state polarization decays according to the decoherence rate $(\Gamma_{\alpha_1} + \Gamma_{\alpha_2})/2$.
In contrast, by inserting into Eq.~(\ref{SBE}) the explicit form of the Lindblad superoperator (\ref{Lindblad}), it is easy to get
\begin{equation*}
\frac{\upd \rho_{\alpha_1\alpha_2}^{\rm i}}{\upd t}
=
\left(
\mathcal{L}_{\alpha_1\alpha_2,\alpha_1\alpha_2}
+
\mathcal{L}_{\alpha_2\alpha_1,\alpha_2\alpha_1}
\right)
\rho_{\alpha_1\alpha_2}^{\rm i}
+ 
\end{equation*}
\begin{equation}\label{SBELindblad}
+ \sum_{\alpha_1'\alpha_2' \ne \alpha_1\alpha_2}
\left(
e^\frac{(\epsilon_{\alpha_1'} - \epsilon_{\alpha_2'} - \epsilon_{\alpha_1} + \epsilon_{\alpha_2}) t}{i \hbar}
\mathcal{L}_{\alpha_1\alpha_2,\alpha_1'\alpha_2'}
\rho_{\alpha_1'\alpha_2'}^{\rm i}\, + \textrm{H.c.}\right)
\end{equation}
with
\begin{equation}\label{calL}
\mathcal{L}_{\alpha_1\alpha_2,\alpha_1'\alpha_2'} = \frac{1}{2} \sum_s
\left(
\mathcal{P}^s_{\alpha_1\alpha_2,\alpha_1'\alpha_2'} 
-\sum_{\alpha'}
\mathcal{P}^{s *}_{\alpha'\alpha',\alpha_1\alpha_1'}
\delta_{\alpha_2\alpha_2'}
\right)\ .
\end{equation} 
In the presence of strongly nonelastic interaction processes, the overall impact of the second term in (\ref{SBELindblad}) is strongly reduced thanks to the fast temporal oscillations of the various free-rotation phase factors; moreover, taking into account that in such nonelastic-interaction limit $\mathcal{P}^s_{\alpha\alpha',\alpha\alpha'} \to 0$, one gets 
$\mathcal{L}_{\alpha\alpha',\alpha\alpha'} \to -\Gamma_{\alpha}/2$, which implies that in this limit the Lindblad-model equation in (\ref{SBELindblad}) reduces to the relaxation-time one in (\ref{SBERTA}). In contrast, in the presence of quasielastic processes one deals with a significant cancelation between in- and out-scattering contributions, not accounted for by the relaxation-time equation (\ref{SBERTA}).
It is worth stressing that such intrinsic limitation of relaxation-time models has been already recognized in the analysis of ultrafast phenomena in photoexcited semiconductors \cite{Rossi02}, showing that the latter becomes particularly severe for the case of quasielastic processes \cite{Rossi94}.

\begin{figure}
\onefigure{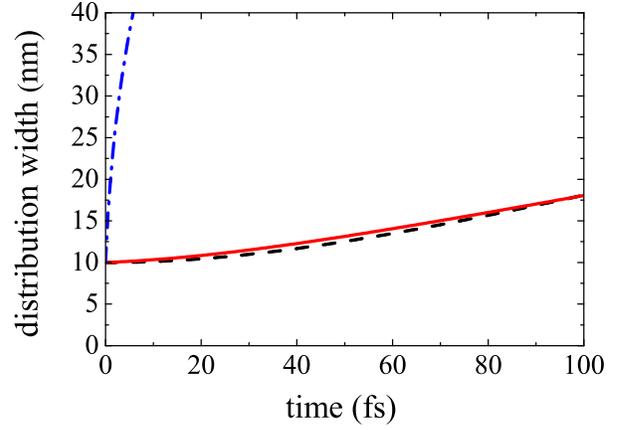}
\caption{
Same as in Fig.~\ref{Fig4} but for a reduced value of the LO-phonon energy ($\epsilon_{\rm LO} = 20$\,meV) (see text).
}
\label{Fig5}       
\end{figure}

To confirm this physical interpretation, we have repeated the simulated experiments presented so far reducing the GaN LO-phonon energy by a factor $4$ (from $80$ to $20$\,meV), such to mimic the quasielastic-process limit.
The time evolution of the effective distribution width $\lambda$ corresponding to these new simulations is reported in Fig.~\ref{Fig5}.
As expected, compared to the results reported in Fig.~\ref{Fig4}, the decoherence overestimation produced by the relaxation-time model (dash-dotted curve) is significantly increased, while the diffusion speed up induced by the Lindblad superoperator (solid curve) is strongly reduced. Indeed, in spite of the fact that the LO-phonon energy is still significantly different from zero, the effect of phonon scattering is already negligible.
This is a clear indication that in the presence of genuine quasi elastic processes like, e.g., carrier-acoustic phonons or carrier-carrier scattering (i) the relaxation-time model is definitely inadequate, and (ii) quantum diffusion due to scattering non-locality is expected to play a minor role.

\begin{figure}
\onefigure{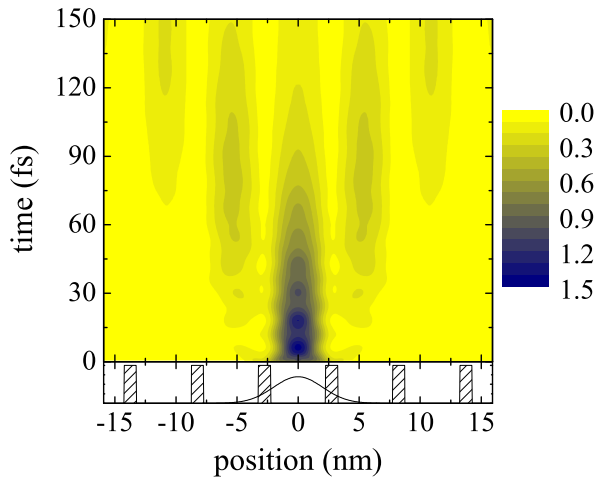}
\caption{
Room-temperature quantum-diffusion dynamics in a GaN-based superlattice (band offset of $0.3$\,eV and well and barrier widths of $4.5$ and $1$\,nm) obtained via the Lindblad scattering superoperator in (\ref{Lindblad}): sub-picosecond time evolution of the spatial carrier density corresponding to the initial mixed state in (\ref{rhocirc}) with $\overline{\Delta}_z = 2$\,nm (see text).
}
\label{Fig6}
\end{figure}

As a final simulated experiment aimed at showing the power and flexibility of the present quantum-transport approach, we have investigated the ultrafast carrier dynamics in a GaN-based superlattice.
Figure \ref{Fig6} displays the sub-picosecond time evolution of the spatial carrier density corresponding to the initial mixed state in (\ref{rhocirc}) with $\overline{\Delta}_z = 2$\,nm, obtained employing the Lindblad scattering superoperator in (\ref{Lindblad}).
Compared to the corresponding bulk-system result of Fig.~\ref{Fig1}, here the presence of the superlattice structure (see lower panel) produces a slow down of the diffusion process (notice that, in order to produce a similar effect, here the initial distribution width has been reduced by a factor 5 (see Eq.~(\ref{Deltazt}))).
Indeed, the semiconductor nanostructure gives rise to a non-trivial interplay between the spatial quantum confinement dictated by the nanostructure potential and the scattering-induced diffusion, resulting in a superlattice-induced modulation of the density profile.
Such space dependent phenomena ---not possible via Boltzmann-like Monte Carlo simulation schemes--- constitutes a distinguished feature of the present quantum mechanical treatment.

Let us finally stress that in the presence of particularly strong interaction mechanisms as well as of extremely short electromagnetic excitations, the application of the Markov limit becomes questionable \cite{Rossi02,Axt04}; however, for a wide range of nanodevices and operation conditions the present Markov treatment is expected to well reproduce the sub-picosecond dynamics induced by a large variety of single-particle scattering mechanisms.

\acknowledgments
We are grateful to Rita Claudia Iotti for stimulating and fruitful discussions.


\begin{thebibliography}{10}
\expandafter\ifx\csname url\endcsname\relax\def\url#1{\texttt{#1}}\fi

\bibitem{b-Jacoboni89}
\Name{Jacoboni C. \and Lugli P.} \Book{The Monte Carlo Method for Semiconductor
  Device Simulation} (Springer) 1989.

\bibitem{Ryzhii98}
\Name{Ryzhii M. \and Ryzhii V.} \REVIEW{Appl. Phys. Lett.}{72}{1998}{842}.

\bibitem{Iotti01}
\Name{Iotti R.~C. \and Rossi F.} \REVIEW{Phys. Rev. Lett.}{87}{2001}{146603}.

\bibitem{Koehler02}
\Name{Kohler R., Tredicucci A., Beltram F., Beere H.~E., Linfield E.~H., Davies
  A.~G., Ritchie D.~A., Iotti R.~C. \and Rossi F.}
  \REVIEW{Nature}{417}{2002}{156}.

\bibitem{Callebaut04}
\Name{Callebaut H., Kumar S., Williams B.~S., Hu Q. \and Reno J.~L.}
  \REVIEW{Appl. Phys. Lett.}{84}{2004}{645}.

\bibitem{Lu06}
\Name{Lu J.~T. \and Cao J.~C.} \REVIEW{Appl. Phys. Lett.}{89}{2006}{211115}.

\bibitem{Bellotti08}
\Name{Bellotti E., Driscoll K., Moustakas T.~D. \and Paiella R.} \REVIEW{Appl.
  Phys. Lett.}{92}{2008}{101112}.

\bibitem{Jirauschek10}
\Name{Jirauschek C.} \REVIEW{Appl. Phys. Lett.}{96}{2010}{011103}.

\bibitem{Matyas10}
\Name{Matyas A., Belkin M.~A., Lugli P. \and Jirauschek C.} \REVIEW{Appl. Phys.
  Lett.}{96}{2010}{201110}.

\bibitem{Iotti10}
\Name{Iotti R.~C., Rossi F., Vitiello M.~S., Scamarcio G., Mahler L. \and
  Tredicucci A.} \REVIEW{Appl. Phys. Lett.}{97}{2010}{033110}.

\bibitem{Matyas13}
\Name{Matyas A., Lugli P. \and Jirauschek C.} \REVIEW{Appl. Phys.
  Lett.}{102}{2013}{011101}.

\bibitem{b-Datta05}
\Name{Datta S.} \Book{Quantum Transport: Atom to Transistor} (Cambridge
  University Press) 2005.

\bibitem{b-Haug04}
\Name{Haug H. \and Koch S.} \Book{Quantum Theory of the Optical and Electronic
  Properties of Semiconductors} (World Scientific) 2004.

\bibitem{b-Haug07}
\Name{Haug H. \and Jauho A.} \Book{Quantum Kinetics in Transport and Optics of
  Semiconductors} (Springer) 2007.

\bibitem{b-Jacoboni10}
\Name{Jacoboni C.} \Book{Theory of Electron Transport in Semiconductors: A
  Pathway from Elementary Physics to Nonequilibrium Green Functions} (Springer)
  2010.

\bibitem{b-Rossi11}
\Name{Rossi F.} \Book{Theory of Semiconductor Quantum Devices: Microscopic
  Modeling and Simulation Strategies} (Springer) 2011.

\bibitem{Gmachl01}
\Name{Gmachl C., Capasso F., Sivco D.~L. \and Cho A.~Y.} \REVIEW{Rep. Prog.
  Phys.}{64}{2001}{1533}.

\bibitem{Iotti05a}
\Name{Iotti R.~C. \and Rossi F.} \REVIEW{Rep. Prog. Phys.}{68}{2005}{2533}.

\bibitem{Rossi02}
\Name{Rossi F. \and Kuhn T.} \REVIEW{Rev. Mod. Phys.}{74}{2002}{895}.

\bibitem{Axt04}
\Name{Axt V.~M. \and Kuhn T.} \REVIEW{Rep. Prog. Phys.}{67}{2004}{433}.

\bibitem{Brunetti94}
\Name{Brunetti R., Jacoboni C. \and Price P.~J.} \REVIEW{Phys. Rev.
  B}{50}{1994}{11872}.

\bibitem{Reiter07}
\Name{Reiter D., Glanemann M., Axt V.~M. \and Kuhn T.} \REVIEW{Phys. Rev.
  B}{75}{2007}{205327}.

\bibitem{Rosati13b}
\Name{Rosati R. \and Rossi F.} \REVIEW{Appl. Phys. Lett.}{103}{2013}{113105}.

\bibitem{Taj09}
\Name{Taj D., Iotti R.~C. \and Rossi F.} \REVIEW{Eur. Phys. J.
  B}{72}{2009}{305}.

\bibitem{Iotti05b}
\Name{Iotti R.~C., Ciancio E. \and Rossi F.} \REVIEW{Phys. Rev.
  B}{72}{2005}{125347}.

\bibitem{Lindblad76}
\Name{Lindblad G.} \REVIEW{Commun. Math. Phys.}{48}{1976}{119}.

\bibitem{Dolcini13}
\Name{Dolcini F., Iotti R.~C. \and Rossi F.} \REVIEW{Phys. Rev.
  B}{88}{2013}{115421}.

\bibitem{Hess96}
\Name{Hess O. \and Kuhn T.} \REVIEW{Phys. Rev. A}{54}{1996}{3347}.

\bibitem{Demeio05}
\Name{Demeio L., Bordone P. \and Jacoboni C.} \REVIEW{Transport Theor.
  Stat.}{34}{2005}{499}.

\bibitem{Rosati13a}
\Name{Rosati R., Dolcini F., Iotti R.~C. \and Rossi F.} \REVIEW{Phys. Rev.
  B}{88}{2013}{035401}.

\bibitem{Gebauer04}
\Name{Gebauer R. \and Car R.} \REVIEW{Phys. Rev. Lett.}{93}{2004}{160404}.

\bibitem{b-Ashcroft11}
\Name{Ashcroft N. \and Mermin N.} \Book{Solid State Physics} (Cengage Learning
  India Private Limited) 2011.

\bibitem{Rossi94}
\Name{Rossi F., Haas S. \and Kuhn T.} \REVIEW{Phys. Rev. Lett.}{72}{1994}{152}.

\end{thebibliography}

\end{document}